\documentclass[twocolumn]{sigcomm-alternate}

\usepackage{graphicx,amsmath,amssymb,url,subfigure}

\renewcommand{\geq}{\geqslant}

\begin{document}

\title{
    On Compact Routing for the Internet
}

\numberofauthors{4}
\author{
    \alignauthor Dmitri Krioukov \\ \affaddr{CAIDA} \\ \email{dima@caida.org}
    \alignauthor kc claffy \\ \affaddr{CAIDA} \\ \email{kc@caida.org} \and
    \alignauthor Kevin Fall \\ \affaddr{Intel Research Berkeley} \\ \email{kfall@intel.com}
    \alignauthor Arthur Brady \\ \affaddr{Tufts University} \\ \email{abrady@cs.tufts.edu}
}

\maketitle

\begin{abstract}

The Internet's routing system is facing stresses due to
its poor fundamental scaling properties.
Compact routing is a research field that
studies fundamental limits of routing scalability
and designs algorithms that try to meet these limits.
In particular, compact routing research shows that
shortest-path routing, forming a core of traditional
routing algorithms, cannot guarantee routing table~(RT)
sizes that on all network topologies grow slower than
linearly as functions of the network size. However,
there are plenty of compact routing schemes that relax the
shortest-path requirement and allow for improved,
sublinear RT size scaling that is mathematically provable for
all static network topologies.
In particular,
there exist compact routing schemes designed for grids, trees,
and Internet-like topologies that offer RT sizes
that scale logarithmically with the network size.

In this paper, we demonstrate that in view of recent results
in compact routing research, such logarithmic scaling on Internet-like topologies
is fundamentally impossible in the presence of
topology dynamics or topology-independent (flat) addressing.
We use analytic arguments to show that the number of routing
control messages per topology change cannot scale better than
linearly on Internet-like topologies. We also employ simulations
to confirm that logarithmic RT size scaling gets broken by
topology-independent addressing, a cornerstone of popular
locator-identifier split proposals aiming at improving
routing scaling in the presence of network topology dynamics
or host mobility. These pessimistic findings lead us to
the conclusion that a fundamental re-examination of assumptions
behind routing models and abstractions is needed in order
to find a routing architecture that would be able to scale
``indefinitely.''

\end{abstract}

\category{\\G.2.2}{Graph Theory}{Graph algorithms, Network problems}
\category{\\C.2.1}{Network Architecture and Design}{Network topology}



\section{Introduction}

Despite prevailing concerns among network operators and developers
that the current Internet interdomain
routing system will not scale to meet the needs of the 21st century
global Internet, networking research has not yet produced
a new routing architecture with satisfactory flexibility and
mathematically provable scalability characteristics.
We believe that we will soon be faced with a critical
architectural inflection point posing a significant
challenge to scaling the size of the future Internet.
The last critical point was reached when the Internet's
routing system adopted strong address aggregation using CIDR
to handle address scaling in the mid 90's. While CIDR was
an extremely effective tactic, most experts agree that
the growth behavior of the routing table in the last decade
confirms that the type of address management required
by CIDR will not suffice to
meet future Internet routing needs, and that a fundamental top-to-bottom
reexamination of the routing architecture may be required~\cite{iab-raws-report}.

In this paper we set to work on such a reexamination.\footnote{We clarify up front
that scalability is only one of several
problems of the current Internet routing architecture. Other
problems include security, isolation,
configuration control, etc. See~\cite{routing-requirements} for a long
list of future routing architecture requirements.}
As we proceed, we first identify, in Section~\ref{sect:scalability},
the fundamental causes of Internet routing scalability problems,
and how they imply the need for
dramatically more efficient routing algorithms with rigorously
proven worst-case performance guarantees.
Such algorithms exist---they are known collectively as {\em compact
routing schemes}. They do not and, generally, can not guarantee
routing along shortest paths---they {\em stretch\/} them. In
Section~\ref{sect:stretch}, we juxtapose the formal notion of
stretch (or {\em hop stretch}) with better-known forms of path
inflation present in conventional networking. There is a fundamental
trade-off between stretch and routing table~(RT) size, and we show
in Section~\ref{sect:noscale} why hierarchical routing and
addressing fall short in finding an optimal balance point of this
trade-off for topologies of interest. We also outline, in the same
section, known alternative addressing techniques that compact
routing schemes use. We analyze the key ideas behind these schemes
in Section~\ref{sect:compact_routing}. We focus on universal schemes
in this section and shift our attention to the schemes specialized
for Internet-like topologies in Section~\ref{sect:applicability}.
All these schemes are static. They ignore any communication overhead
their implementations might require.
We use
analytic arguments in Section~\ref{sec:korman-peleg} to directly
prove that recent results in compact routing research render
fundamentally unscalable the communication costs of any routing
algorithm applied to Internet-like topologies. Despite these findings,
``locator-identifier split''~(LIS) proposals are
becoming popular these days, as they are widely believed to scale
better in the presence of network topology dynamics and host mobility.
In Section~\ref{sec:dynamic} we establish that LIS is a form of
name-independent compact routing, discussed in
Section~\ref{sect:compact_routing} and using topology-unaware (flat)
addressing. We show that contrary to common belief, the scaling
properties of such routing can only be worse compared to the
topology-aware, name-dependent case. We employ simulations to
demonstrate that both the RT sizes and stretch induced by the
best-performing name-independent schemes are significantly worse
compared to their name-dependent counterparts. We conclude and pose
open problems in Section~\ref{sect:conlcusion}.

\section{Internet Routing Today}
\label{sect:scalability}

There are two main related problems with the Internet's current
routing architecture.
First, it is not flexible, due primarily to the need to assign
node addresses based on topological location.
In order to achieve highly efficient address aggregation
and relatively small routing tables, a CIDR network must have
two characteristics: an essentially tree-like graph structure,
and address assignment that follows this structure.
Node mobility and end site multi-homing fundamentally break this model.
But even recommended Internet traffic engineering techniques,  e.g.,
load balancing, require {\em de-}aggregation of address blocks
that cause the routing table to grow even if the network
itself is not growing.

The second problem with the Internet's routing architecture
is how the interdomain routing algorithm (path vector, for BGP)
and its protocol implementation scale with the size of the network.
Poor scaling of a routing algorithm expresses itself in terms
of rapid, e.g., linear, rates of growth of the routing table~(RT) size.
Poor scaling of RT sizes exacerbates
convergence problems (convergence time, churn,
instabilities, etc.).
Not only is the communication overhead of BGP
known to be exponential~\cite{LaAhBo01}, but the BGP RT size also
appears to grow exponentially~\cite{huston-bgp-site}.

The immediate causes of Internet RT growth have been extensively
analyzed~\cite{iab-raws-report,huston01-03,huston01-02,BroNeCla02,NaGoVa03}; they
reflect voluntary business relationships occurring at increasing
density, traffic engineering techniques to improve robustness
or reduce operational costs, and address allocation policies
developed by IANA and the regional address registries---all resulting
in various forms of de-aggregation, which is
the primary cause of {\em super-linear\/} growth of the BGP RT size.
Several studies, proposals, and even new routing architectures seek to
mitigate these effects. Prominent efforts in this area
include~\cite{beyond-bgp,atoms,multi6,islay,nimrod} and most
recently~\cite{GuKoKi04,SuKaEe05,CaCo06}.
However, these approaches do not rigorously provide non-trivial scaling guarantees,
and many of them represent only short-term fixes since they address the
symptoms rather than the root of the problem.
More precisely, most of these approaches try to find shortest paths in the
topology, leading to RT sizes that cannot scale better than linearly~\cite{GaPe96}.
We are interested in algorithms that scale better than that.
Ideally, we want to achieve logarithmic scaling\footnote{To be more precise, we mean
{\em polylogarithmic\/} scaling, but we abuse the terminology and do not
differentiate between logarithmic and polylogarithmic growth in this paper.}
that comes close to satisfying our desire for ``infinite scalability''
in a future routing architecture~\cite{routing-requirements}.

As a representative example of the efforts mentioned above, we consider the idea of
routing on autonomous system~(AS) numbers, wherein AS numbers
replace IP prefixes in the role of interdomain routing addresses.
Using the coarser granularity of AS numbers
is a natural idea, and is now a common part of many
proposals~\cite{atoms,islay,nimrod,GuKoKi04,SuKaEe05}.
At first glance, routing on AS numbers looks like a
solution to the RT size problem.  Indeed, such an approach
could immediately reduce the RT size by an order of magnitude,
since there are on the order of $10^5$ IP prefixes and $10^4$ AS numbers
in the global RT today.  However, one can easily overlook that it is just a one-time
constant-factor reduction, and not a change of the scaling behavior.
As such, it is not a solution to the core problem, but rather a temporary
relief measure.
Indeed, if algorithms routing on AS numbers
are still shortest-path routing algorithms, then they still cannot produce RT sizes that grow
slower than linearly with the network size measured, in this case, by the total
number of ASs in the Internet. To make things even worse,
this number currently
grows faster than the total number of IP
prefixes~\cite{huston-bgp-site,BroNeCla02}, and if the trend continues after
a hypothetical adoption of routing on AS numbers, then the actual {\em rate of growth\/}
of RT sizes can only be worse than today.
Furthermore, routing on AS numbers requires some form of global directory
to map IP addresses or prefixes
to the AS numbers they are associated with.

So far we have argued that inflexibility of the current
routing architecture precludes long-term scalability.
An ideal routing architecture would allow for
assigning node addresses independent
of topology, and would employ routing algorithms that can truly scale.
No proposal thus far has simultaneously solved both objectives;
address aggregation remains the only generally accepted method for
effectively limiting RT size growth, yet operational and business
imperatives require intentional violation of this tenet.
Something has to give.

In order to find sustainable
solutions to the global Internet routing problem, we need to
investigate routing scalability at its most fundamental level.
For interdomain routing, the framing parameters are the performance
guarantees of routing algorithms operating on graphs with
topologies similar to the Internet AS-level topology, and
which can support topology-independent addressing of nodes.  In particular,
we seek rigorous lower and upper bounds on the RT size used in supporting
all routing features, i.e., mapping node names to their locations,
and reaching those locations.
Compact routing research establishes these bounds and constructs algorithms
(or schemes) that try to meet them. One of the central themes in compact routing
is the investigation of the inevitable trade-off between RT sizes and routing stretch.\\

\section{Routing stretch}
\label{sect:stretch}

The {\em stretch} of a routing algorithm is defined as the
worst-case path-length increase factor relative to shortest paths.
More specifically, for every pair of nodes in all graphs
in the set of graphs the algorithm can operate on,
we find the ratio of the length of the route taken by the
algorithm to the length of the shortest available path between the same pair
of nodes. The maximum of this ratio among all node pairs in all
the graphs in the set is the algorithm's stretch.
The average stretch is the average of this ratio across all source-destination
pairs in a given graph or set of graphs.

We emphasize that this notion of stretch has
nothing in common with known forms of ``path inflation''
in contemporary Internet routing.
Path inflation in today's Internet is {\em not\/} due to the
stretch of the underlying routing algorithm, but rather due to
intra- and inter-domain routing policies, and various incongruities
across multiple levels (geographical, router, AS, etc.)
of network topology abstraction~\cite{SprMaAn03}.

We are not aware of any widely-used routing protocol
that produces other than stretch-1 (shortest) paths by default.
Link-state~(LS) in OSPF or ISIS, distance-vector~(DV) in RIP or (E)IGRP,
LS/DV hybrid in LVA~\cite{BeGa04}, and path-vector~(PV) in BGP,
are all forms of
{\em trivial}
shortest-path routing (stretch-1).  Consider two examples:
BGP and OSPF.   Non-trivial policy configurations generally
prevent BGP from routing along the actual shortest path,
but without policies the BGP route selection process
would always select the shortest paths in the AS-level graph.
Similarly, non-trivial area configurations prevent OSPF
from routing along the shortest paths, but routing inside
an area is always along the shortest path in the weighted router-level graph.

The above examples seem straightforward, but we identify them
to eliminate any possible confusion between stretch of a
routing {\em algorithm\/} and inflation of paths produced by a routing
{\em protocol} subjected to policy or configuration constraints.
Clarifying the difference allows us to formulate
both necessary and sufficient conditions for the stretch of a
routing algorithm proposed for use in the Internet.

The {\em sufficient} routing stretch requirement is that
the routing algorithm underlying
any realistic interdomain routing protocol must be of stretch-1.
To understand the reasoning behind this requirement,
consider a
stretch$>$1 routing algorithm and
apply it to a complete graph, which has all nodes directly
connected to all other nodes so that all shortest paths are of length~1.
Since this algorithm does not guarantee finding all shortest paths,
RTs at some nodes might lack entries corresponding to
their own directly-connected neighbors. Put in different terms,
in order to shrink RTs, the algorithm might have to force
some nodes to delete information about some of their
directly-attached interfaces.

Of course, the Internet interdomain topology is not a full mesh,
and it might turn out that the same routing algorithm that fails
to be stretch-1 on a full mesh does in fact find all shortest paths
to each node's neighbors in realistic Internet-like topologies.
We thus might also pursue the explicit task of finding
a non-universal routing algorithm that is applicable only
to Internet-like topologies.
We do not require this algorithm be stretch-1, but it
must be of stretch-1 on paths of length~1 leading to nodes' neighbors.
In other words, the {\em necessary} routing stretch requirement
is that the routing algorithm underlying any realistic
interdomain routing protocol must be of stretch-1 {\em on paths
of length~1} in realistic Internet-like topologies.

These requirements are purely practical: if a link exists between
a pair of autonomous systems,
it must be made available for carrying traffic.
If our ``sophisticated'' routing algorithm removes routing
information about such links from routing tables, then network
operators will feel compelled to manually reinsert it, which counters
the scalability objective of reducing the RT size and requires
manual intervention which we would in general like to avoid.
We will refer to this (undesirable) administrative re-adding
of information about shortest paths to nodes' neighbors
as {\em manual reinsertion}.

The absence of an immediate neighbor in a node's RT appears
to contradict common sense, but one can verify that this situation
can occur with the algorithms we discuss in
Section~\ref{sect:compact_routing}.   We have not had to deal with
this effect in practice yet
because the routing algorithms in operation today are all stretch-1.
Stretch-1 algorithms require complete knowledge of the
network topology (or at least of distances between nodes in it)
and maintenance of an individualized optimal next
hop for every possible destination in the network.
This means $\Omega(n)$ entries are required in the RTs of every node in the worst case.
In order to shrink RTs below this size,
we must prepare to {\em lose} topological information, and consequently
lose optimal next hops for some destinations.
Balancing this fundamental and unavoidable
trade-off between stretch of a
routing algorithm and the sizes of RTs it produces is at the center of
compact routing research today, but the first formalization of
this trade-off was introduced in early work on
hierarchical routing and addressing.

\section{Hierarchical routing, addressing, and their alternatives}
\label{sect:noscale}

In 1977 Kleinrock and Kamoun published
their pioneering paper on hierarchical routing~\cite{KK77}.
They showed how hierarchical clustering could be used with
appropriate node addresses to produce highly scalable routing tables.
This technique is the basis of the popular {\em hierarchical routing}
approaches used today in practice, e.g., in the form of CIDR in the
interdomain routing case, or usage of OSPF/ISIS areas in the intradomain
routing case.
The key idea is to group (or aggregate)
nearby nodes into clusters, clusters into super-clusters, and so on,
in a bottom-up fashion.
Determining these clusters can be accomplished through
a {\em hierarchical clustering\/} algorithm, the performance of which
is key to good routing performance.

Hierarchical aggregation and addressing offer
substantial RT size reduction by abstracting out
unnecessary topological details about
remote portions of the network: nodes in one \mbox{(super-)cluster} need to keep
only one RT entry for all nodes in another \mbox{(super-)cluster.}
Almost all proposals for future Internet routing
architectures trying to address the RT size problem are based,
explicitly or implicitly, on this concept of hierarchical routing. In
fact, no other concept behind RT size reduction has
been systematically considered in the networking literature.
Unfortunately, hierarchical routing is simply not a good candidate for
interdomain routing, despite its current widespread use and
reputation as realistically scalable, and we show why in the rest of this section.

In the same paper~\cite{KK77}, Kleinrock and Kamoun were the
first to analyze the stretch/RT size trade-off. They showed that the
routing stretch produced by the hierarchical approach
is satisfactory only for
topologies with average shortest path hop-length (distance)~$\bar{d}(n)$
that grows quickly with network size~$n$, which means
it performs well only for graphs where large distances between nodes prevail.
In fact, they assumed that the average distance grows polynomially
with the network size, \mbox{$\bar{d}(n) \sim n^\nu$} with~$\nu>0$.
Such graphs have many remote nodes, i.e., nodes at long hop distances
from each other, which allows for efficient aggregation of topology
details without substantial path length increase.

\subsection{Grids}\label{sec:grids}

The simplest example of graphs with average
distance \mbox{$\bar{d}(n) \sim n^\nu$} are $\delta$-dimensional
grids. The exponent~$\nu$ is~$1/\delta$ in this case, and node
aggregation works naturally: clusters are composed from nodes located
at small Manhattan distances from each other, while super-clusters
contain neighboring clusters, and so on.

We emphasize that
aggregation is not the only technique to achieve logarithmic RT scaling
for grids.
We could alternatively embed a $\delta$-dimensional grid into a
$\delta$-dimensional Euclidean space, and address nodes by their
coordinates in the space.  With such an embedding, nodes would not
need to keep any RT information except for their own addresses,
since knowing their own coordinates and the coordinates of the
destination allows for shortest-path routing according to the
Euclidean, i.e., Manhattan, distance.  The RT size thus scales as the
address size, i.e., logarithmically.

Logarithmic scaling extends from
$\delta$-dimensional grids to graphs with
{\em doubling dimension}~$\delta$~\cite{AbGa06}. The doubling dimension of
a graph, or more generally, of a metric space, is the smallest
$\delta$ such that every ball of radius~$2r$ can be covered by
$2^\delta$ balls of radius~$r$. Informally, such graphs are
``perturbed versions'' of $\delta$-dimensional grids. A key
idea behind efficient addressing in such graphs allowing for logarithmic
RT size growth is to introduce distance scales~\cite{InMa04,AbBa05} or
maps of node coordinates. The address of a node becomes a collection of
its coordinates, which are the distances from the node
to carefully constructed subsets of nodes that form a ``frame of
reference.'' In other words, knowing the distances to these frame-of-reference
subsets, we can uniquely identify a node's location in the graph.
One can show then that in order to bring stretch arbitrarily close to~1,
we need only a logarithmic number of such subsets, which guarantees
logarithmic RT sizes.

\subsection{Trees}\label{sec:trees}

A strikingly similar situation occurs with routing on regular, i.e., $b$-ary, trees.
They also allow for efficient aggregation: combine leaves of a
single parent into a single cluster, while super-clusters contain clusters
with a common parent of the parent, and so on.

Yet again we find that aggregation is not the only path toward logarithmic
RT scaling on such trees.  For example, we can use node addressing induced
by a depth-first-search (DFS) node numbering.
Given a DFS-based addressing convention, any node can
infer its exact location on the tree from its address.  In
particular, a node knows what nodes are its children or descendants of
which one of its children, and what nodes are not its descendants.

As with grids, logarithmic scaling extends from regular $b$-ary trees
to a wider class of similar graphs---to arbitrary trees in this
case~\cite{FraGa01,ThoZwi01b}. The main idea behind efficient addressing
schemes is still to use DFS numbering, but we can no longer rely only on
the node DFS ID to identify its location on a tree. It turns out however
that it is possible to encode the path from the root of a tree
to a node in such a way that this encoding preserves the logarithmic
upper bound. If we now make the sequence of nodes lying on the path to
the root a part of the node address, then one can check that routing
along shortest paths becomes trivially possible.\\

\subsection{Grids and trees}\label{sec:grids-trees}

In both trees and grids,
shortest-path routing is possible given only
the addresses of the current and destination nodes.
The RT size in either case scales logarithmically. This scaling performance
can be extended beyond regular grids and trees to much wider classes of
graphs, but not with hierarchical routing.
The simplest example is routing on star graphs.
These graphs are trees, but
no hierarchical aggregation is possible on them. Logarithmic scaling
in such cases is possible due only to specialized addressing
techniques that leverage topological peculiarities of a graph's structure
to succinctly encapsulate the location of a node into its address.
Of course, these addressing techniques work trivially for
regular grids and trees as well.

Recent progress in this area indicates that the
efficiency of these types of addressing schemes is directly
related to existence of low-distortion embeddings of finite metric
spaces induced by shortest-path distances in grids and trees
into normed spaces of low dimension~\cite{matousek02}. Low dimensions
of host spaces guarantee small RT sizes, while low distortion of
embeddings corresponds to small stretch. Grids embed in flat Euclidean
spaces, but trees require spaces of negative
curvature~\cite{KraLe06,kleinberg07infocom}.

\subsection{Scale-free graphs}\label{sec:scale-free}

According to the best available data~\cite{MaKrFo06}, the Internet
topology is neither a tree nor a grid.
Rather, it is {\em scale-free}.
To avoid terminology disputes, by {\em scale-free\/}
we simply mean networks with heavy-tail, e.g., power-law,
node degree distributions and strong clustering, i.e., large
numbers of triangular subgraphs.\footnote{There is an unfortunate
terminological collision here: {\it clustering} means both
grouping nodes into areas and presence of triangles in a graph.}
The latter characteristic
implies that such graphs are not trees at all. The former,
on the other hand, means that in theory, the average
distance in them
grows with the network size
much more slowly than Kleinrock and Kamoun~\cite{KK77} assumed:
at most \mbox{$\bar{d}(n) \sim \log n$}~\cite{ChLu03}.  In practice,
the average hop distance between autonomous systems in the
Internet stays virtually
constant or even decreases due to increasing inter-AS connectivity
driven by economic or performance considerations~\cite{LeCleFa05,
BroNeCla02,huston-bgp-site}.  According to multiple data sources,
the average AS-hop distance in the current Internet is between~3.1 and~3.7,
with more than~80\% of AS pairs being \mbox{2-4} hops away from each
other~\cite{MaKrFo06}.
Topologies with small average distances are also called
{\em small worlds}.
There are
essentially {\em no} remote nodes in small-world networks;
all nodes are close to each other.

These characteristics are extremely bad news for hierarchical routing approaches,
because the effectiveness of hierarchical network partitioning
and aggregation
depends either on the abundance
of remote notes or on strong regularity of $b$-ary tree structure.
None of these properties is present in scale-free graphs.
In short, we are faced with an unsettling reality: {\em one cannot hope
to find a way to efficiently apply hierarchical, aggregation-based
routing to Internet-like topologies.}

These general arguments do not preclude the possibility of applying
{\em some} generic hierarchical routing algorithm to the Internet.
But they do imply that such an application cannot be efficient and we
must prepare to see high stretch. Simple analytical estimates~\cite{KrFaYa04}
show that applying hierarchical routing to an Internet AS-level
topology incurs a $\sim$15-times path length increase, which,
although alarming enough by itself, would also lead to a substantial
RT size surge caused by manual reinsertion.

If we accept our argument in Section~\ref{sect:stretch}
that any truly scalable Internet routing algorithm must have
stretch as close as possible to~1, we must also accept the
fact that hierarchical routing will not meet our needs.
Since all previous Internet interdomain routing proposals are
heavily based on hierarchical routing,
we recognize that while deploying one of them might offer
short-term relief, none of them would genuinely scale to
meet future demands.

\section{Universal compact routing}
\label{sect:compact_routing}

Compact routing research
establishes fundamental limits for routing scalability, and constructs concrete
routing algorithms (or {\it schemes})
that try to meet these limits.
Formally, a routing scheme is said to be {\it compact} if it produces
logarithmic address and header sizes, a
sublinear RT size, and stretch bounded by a constant, as opposed
to the growing stretch more typical of hierarchical routing.\footnote{For
example, one can verify that on an $n$-node full mesh,
\cite{KK77}~produces stretch growing to infinity as~$\Theta(\log n)$.}
A routing scheme is {\it universal} if it works correctly and
satisfies promised scaling bounds on all graphs. If it does so only on
some specific graph classes, it is called {\it specialized}.

We can express fundamental limits of routing scalability using
{\em lower bounds\/} on RT size versus stretch.
For example, we can fix the maximum stretch and
establish the minimum RT size for this stretch.
Such a task usually involves constructing a concrete worst-case graph,
enumerating RT size requirements of {\em all possible\/} routing schemes,
and demonstrating that no routing scheme, applied to this
worst-case graph, produces an RT smaller than the minimum~\cite{GaGe01}.
As soon as we have constructed a concrete routing scheme, we can establish
its {\em upper bounds}, i.e., the maximum RT size and stretch it produces
across all graphs. If a scheme's upper bound equals the theoretical lower bound,
the scheme is said to be {\it optimal}.

The lower bounds for universal \mbox{stretch-1} (shortest-path) routing
are somewhat pessimistic. We first notice that we
can construct RTs at each node
by storing, for every destination node, the ID of the
outgoing port on the shortest path to the destination.
The number of destinations is \mbox{$n-1$}, the maximum number of ports
a node can have (equal to maximum possible node degree)
is also \mbox{$n-1$}, and thus a maximum $O(n \log n)$
bits of memory is required for an RT, i.e., $O(n \log n)$ is its
upper bound. This
construction is called {\em trivial shortest-path routing}, and
all deployed LS-, DV-, or PV-based routing protocols implement it.
However, Gavoille and P\'{e}renn\`{e}s~\cite{GaPe96}
showed that the lower bound of universal stretch-1
routing is also $\Omega(n \log n)$, i.e., there is no shortest
path routing scheme that for all nodes in all graphs, guarantees
RTs smaller than the RTs of trivial shortest-path routing.
In other words, {\em shortest-path routing is incompressible}.
In order to decrease the maximum RT size, we therefore
must allow maximum stretch to increase above~1.

Gavoille and Genegler~\cite{GaGe01} showed that
for any stretch strictly below~3, the local space
lower bound is nearly the same as for stretch-1 routing,~$\Omega(n)$,
meaning that {\em no\/} universal stretch$<$3 routing scheme can
guarantee sublinear RT sizes: the minimum value of maximum stretch
allowing sublinear RT sizes is~3. Thorup and Zwick~\cite{ThoZwi01a}
proved that any routing scheme with stretch strictly
below~5 cannot guarantee space smaller than~$\Omega(n^{1/2})$.
It is widely believed (although not yet proven) that stretch below
\mbox{$2k-1$} imposes an RT size lower bound of~$\Omega(n^{1/k})$
for all positive integers~$k$.

Besides being universal or specialized, another important classification
of routing schemes is whether they support {\em name independence}.
Name-{\em dependent\/} schemes
embed some topological information in node addresses (or {\it labels})
which thus cannot be arbitrary. We analyze a few classic examples of such
addressing in Section~\ref{sect:noscale}.
Name-{\em independent\/}
schemes work on graphs with arbitrary, e.g., flat, node labels.
As discussed in Section~\ref{sect:scalability}, scalable name-independent
schemes are highly desirable for future Internet routing.
The distinction between these two cases is directly related to a popular
discussion theme in the networking community, in which terms such as
``node name'' or ``node identifier'' essentially refer to the name-independent,
topologically agnostic label, while terms ``node address'' or ``node locator''
usually imply a topologically informative node label, i.e., the name-dependent label.
Somewhat surprisingly, the lower and upper bounds for both name-dependent
and name-independent universal routing are essentially the same.

\subsection{Name-dependent routing}

Kleinrock and Kamoun's hierarchical approach~\cite{KK77}
was essentially the first name-dependent routing scheme.
Besides problems discussed in Section~\ref{sect:noscale}, their scheme
provided no algorithm to construct a required partitioning for a
given network. In subsequent works, first Kleinrock himself,
then Perlman, and more recently Awerbuch, Peleg and others,
invested much effort trying to find efficient network partitioning
algorithms needed for correct operation of hierarchical routing.

It became evident that hierarchical routing approaches were
intrinsically suboptimal when Cowen
delivered her compact routing scheme~\cite{cowen01}. The scheme
required no network partitioning, i.e., it was
non-hierarchical, and it was fairly simple compared to numerous
hierarchical routing algorithms accumulated by that time.
It guaranteed the maximum stretch 3 and RT size of $\tilde{O}(n^{2/3})$.
Thorup and Zwick \cite{ThoZwi01b} (TZ) soon improved on Cowen's RT size
upper bound, bringing it to~$\tilde{O}(n^{1/2})$ while maintaining stretch of~3.
The TZ scheme was thus the first nearly optimal universal stretch-3 routing
scheme because its RT size
upper bound was nearly equal to the existing lower bound, up to logarithmic
factors implied by~\mbox{` $\tilde{}$ ' in} the $\tilde{O}$ notation.

The TZ scheme is heavily based on the Cowen scheme.
Both schemes first preprocess the graph and select a
set of landmarks in it. In the Cowen scheme, this selection is based
on results for dominating set construction, while the TZ scheme
selects landmarks by means of randomized techniques.
The landmark set sizes in the Cowen and
TZ cases are $\tilde{O}(n^{2/3})$ and $\tilde{O}(n^{1/2})$, and that
is essentially the only difference between the two schemes. The most
non-trivial part in both the algorithms is to guarantee a proper
balance between the sizes of landmark sets and clusters. Clusters are
defined for every node~$v$ in a graph, and $v$'s cluster is the set
of nodes that are closer to~$v$ than to their corresponding closest
landmarks. The TZ scheme, for example, guarantees that not only the
landmark set size, but also the cluster size is upper-bounded
by~$\tilde{O}(n^{1/2})$. The schemes then assign to all nodes~$v$
the addresses, or labels, consisting of the following three parts:
$v$'s original identifier, the identifier of the landmark~$L(v)$
closest to~$v$, and the identifier of the interface at $L(v)$ that
lies on the shortest path from~$L(v)$ to~$v$. The RT at
$v$ consists of the next hops along shortest paths to all landmarks and all nodes in
$v$'s cluster. The RT size is thus~$\tilde{O}(n^{1/2})$. Node~$v$ forwards a
packet along the shortest path if the destination is either a
landmark or a node in $v$'s cluster. If $v$ is the closest landmark
for the destination, $v$ forwards the packet along the shortest
path using the interface identifier in the address in the packet
header. Otherwise, $v$ forwards the packet to the landmark closest
to the destination---its identifier is in the packet header and
the next hop on the shortest-path route to it is in $v$'s RT. It is a
trivial exercise to check that the worst-case stretch experienced by
such a routing scheme is~3.

\subsection{Name-independent routing}\label{sec:nicr}

Abraham {\em et al.}~\cite{AbGaMaNiTho04} improved on previous
results by Arias {\em et al.}~\cite{ArCoLaRaTa03}
and delivered the first nearly optimal universal name-independent
routing scheme, which had the same upper bounds as the name-dependent
TZ scheme---maximum stretch of~3 and $\tilde{O}(n^{1/2})$ RT sizes.

All name-independent schemes known to us consist of the following two parts:
\begin{itemize}
\item a {\em name-dependent compact routing scheme\/} operating underneath
the name-independent one; and
\item {\em dictionary tables\/} forming an efficiently-distributed\\
database containing information necessary to translate nodes' name-independent
addresses (flat identifiers) to name-dependent addresses (locators).
\end{itemize}
An important property of the dictionary tables is that
they must be ``everywhere dense'' in the topology, so that any
node~$v$ can resolve all identifiers to locators by consulting
dictionary tables of nodes located in $v$'s neighborhood only.
Obviously, node~$v$ has to keep all routes to nodes in its neighborhood.
Routing proceeds as follows: upon receiving a packet destined
to identifier~$X$, node~$v$ forwards a packet to node~$u$ in its
neighborhood that according to the dictionary table construction
rules (dependent on the routing scheme), contains an
identifier-to-locator mapping for~$X$. Upon reaching~$u$, the packet
learns the locator~$x$ for~$X$ and proceeds to~$x$ as dictated by
the underlying name-dependent scheme.

We emphasize that identifier-to-locator mapping is an integral part
of the routing process, and all performance bounds
apply to the combined process of mapping and forwarding.
Specifically, name-independent RTs include both the RTs of underlying
name-dependent schemes and the dictionary tables, whereas total stretch accounts
for travel to a name-resolution node and then from it to the destination along
a name-dependent, potentially stretched route.
This is in contrast to the separation of DNS name resolution and IP routing
in the Internet,
where DNS is essentially an application running on top of IP and requiring
routing to work in the first place.
It is therefore remarkable
that there exist both name-dependent and name-independent universal stretch-3
compact routing schemes with the same $\tilde{O}(n^{1/2})$ RT size upper bound.\\\\\\

\section{Compact routing on scale-free graphs}
\label{sect:applicability}

One might be alarmed by the previous section, as it seems to
conflict with Section~\ref{sect:stretch} where we show that because
of practical reasons, stretch must be close to~1.
The combination of shortest-path routing incompressibility and the
impossibility of \mbox{stretch$<$3} routing with sublinear RT sizes
seem to imply that scalable interdomain routing is impossible. Recall,
however, that the stretch values mentioned in
Section~\ref{sect:compact_routing} refer to the {\em maximum\/}
stretch across all paths in all graphs, meaning that there exists
some worst-case graph\footnote{See~\cite{GaGe01} for an example of
these worst-case graphs.} on which this maximum is achieved. More
importantly, all schemes mentioned in
Section~\ref{sect:compact_routing} are universal. Since they can
work on all graphs, the worst-case graph may not be, and usually is
not, ``Internet-like.'' We can consequently expect the {\em
average\/} stretch of these schemes on scale-free graphs be lower
than~3. But how close is it to~1?

At first glance, it seems unlikely that it could be too close to~1,
since we saw in Section~\ref{sec:scale-free} that the stretch of
hierarchical routing is high on Internet-like graphs,
and there is no apparent reason to believe that compact
routing would be drastically better. In \cite{KrFaYa04} however, the
authors showed that the average performance of the first optimal
stretch-3 scheme~(TZ)~\cite{ThoZwi01b} on scale-free topologies is
much better than its worst case. While TZ's upper bounds are
around~2200 for RT size and~3 for stretch, the average RT size was
found to be around~50 entries for the whole AS-level Internet, and
the average stretch was only~1.1!
These numbers are striking evidence that, on the same topology and
under the same assumptions,
compact routing has dramatically superior scaling characteristics
compared to hierarchical routing~\cite{KK77},
with its stretch of~15 and RTs of unbounded size,
and compared to simple routing on AS numbers (e.g.,~\cite{SuKaEe05}),
with RT sizes of the order of~$10^4$.

The observation in~\cite{KrFaYa04} that scale-free graphs yield
essentially the {\em best possible\/} performance of the TZ scheme
compared to all other graphs was the first indication that
scale-free graphs are optimally structured for high-performance
routing. This indication inspired several research groups to work on
construction of specialized routing schemes~\cite{BraCow06,CaCoDo06}
designed to utilize structural peculiarities of scale-free graphs in
order to improve routing performance guarantees. The schemes
in~\cite{BraCow06,CaCoDo06} achieve logarithmic RT scaling and
infinitesimally small stretch on scale-free graphs. Both schemes are
based on the observation that although scale-free graphs are not
trees,
trees cover scale-free graphs without incurring much stretch.

As a representative example, the scheme by Brady and
Cowen~\cite{BraCow06} (``BC scheme'') grows one shortest-path tree
rooted at the highest-degree node and a small fixed number of
additional trees to cover edges in the fringe of a power-law graph.
Routing is then confined to this tree collection according to the
schemes discussed in Section~\ref{sec:trees}. Since their RT sizes
scale logarithmically, the BC scheme achieves logarithmic scaling.
Specifically, it guarantees a maximum RT size of~$O(\log^2 n)$.  The
$O(\log^2 n)$ scaling means that even if the graph size is
$2^{128}$, i.e., if all IPv6-capable devices independently
participate in global routing, then the maximum routing table size
in such an Internet would still contain not more than~\mbox{$128^2
\sim 16,000$} entries! The corresponding limit for IPv4 is~$\sim
1,000$.

We thus see that even though hierarchical routing approaches based on
address aggregation cannot work efficiently on Internet-like
topologies (Section~\ref{sect:noscale}), more modern routing
algorithms achieve essentially ``infinite'' scalability on
scale-free graphs, if they are {\em static}.
Can there exist any routing schemes that scale as well on
{\em dynamic} networks?

\section{Dynamic routing on scale-free graphs}\label{sec:korman-peleg}

One of the most limiting assumptions behind the routing
algorithms we have considered thus far
is that they assume that the network is effectively static.
More precisely, it is not that the network is actually static {\em per se},
but that the routing algorithms require a full view of the graph
representing the network topology at any given instant in time.
Any topology
change yields a new graph, on which the algorithm might have to
perform its calculations from scratch.  This behavior is characteristic
of currently-deployed Internet routing algorithms as well.

The most basic parameter measuring the
dynamic performance of routing algorithms is the {\em communication
cost}---the number of control messages needed to converge
after a topology change.
Communication costs are as important as RT size or stretch.
They are associated
with nodes requiring properly updated information about the network
topology in order to route.
Deployed routing protocols, for example, require all nodes
to have a coherent {\em full\/} view of
the network topology or of the distances in it. Indeed,
link-state algorithms, typically used for intra-domain routing,
require each node to have a complete view of {\em all\/} links in
the network, and each node performs a shortest-path computation on
essentially a static graph.  Distance- or path-vector algorithms
need each node to know distances or paths to {\em all\/} other nodes
in the network.

In order to achieve such coherent full views of the
network topology, timely
update messages, a.k.a.\ {\em routing updates}, are employed.
Updates requiring recalculation of routing tables can
lead to delay, instabilities, churn, and other
complications. For example,
there are BGP oscillation scenarios with infinite communication
costs~\cite{GriWil99,GriWil02}, and even during more typical BGP
behavior, they are exponential~\cite{LaAhBo01}. High communication
costs and resulting long convergence times are both a critical
problem and an absolutely inevitable implication of growth of the
current routing architecture~\cite{iab-raws-report}.

Using fairly general arguments, Afek {\it et al.}~\cite{AfGaRi89}
showed that the communication cost cannot scale better
than~$\Omega(n)$. Their results applied to universal schemes, so we
could still hope that schemes specialized for scale-free graphs
would behave better, as the $\Omega(n)$ bound could be attained only
on some marginal worst-case graphs. After seventeen years of
little progress on this problem, Korman and Peleg have
recently tried to improve this pessimistic lower bound
in~\cite{KoPe06}. Unfortunately for Internet routing, their results
are even more pessimistic than Afek's since we use them, combined
with simple analytic estimates, to demonstrate in the Appendix that
the communication cost lower bound for scale-free graphs
is~$\tilde{\Omega}(n)$. This finding means that as far as
communication costs are concerned, {\em Internet-like graphs belong
to the class of worst-case graphs, across all possible network
topologies!}\\

\section{Locator-identifier split~(LIS)}\label{sec:dynamic}

Despite all the findings mentioned above,
a popular proposed approach to Internet routing scalability and
flexibility is to separate the identifier and locator
of a node in the network.
In an architecture where one label
identifies a node and a different label indicates its location,
topology changes will only change the locators, which are assumed to
follow topology and allow for aggressive aggregation. Other
functions (access control lists, application and transport layer
identification of endpoints) can use the identifiers.
This would support significantly enhanced flexibility of the Internet
routing system, but does it really address the scalability concerns?

In light of the discussion in Sections~\ref{sec:scale-free}, \ref{sec:nicr},
and~\ref{sec:korman-peleg},
we must admit that LIS cannot really improve routing scalability, for
at least two reasons:
\begin{itemize}
\item Since LIS approaches are based, explicitly or implicitly,
on aggressive aggregation of locators, they cannot improve scaling
of RT sizes because such
aggregation is impossible on scale-free topologies;
\item Even if LIS approaches did not use aggressive aggregation,
they still could not improve scaling, because in addition to
maintaining and updating RTs of locators as before, the networks
must also maintain and update a (distributed) database of identifier-to-locator
mappings.
\end{itemize}

To ground these claims more firmly, we first observe that
name-independent routing from Section~\ref{sec:nicr} naturally
implements LIS. In fact, the only conceptual difference between
name-independent routing and LIS is that the former is more flexible
than the latter, since it does not assume that locators, i.e.,
name-dependent addresses, are amenable to aggressive aggregation.
It is obvious, however, that the
{\em average\/} scaling characteristics of name-independent schemes
cannot be better than name-dependent ones, since the name independent
schemes are essentially name dependent schemes
{\em plus something else}, i.e., plus dictionary tables and
identifier-to-locator name-resolution, which incurs both RT size
increase and stretch, cf.~Section~\ref{sec:nicr}.

\begin{figure*}[t]
    \centerline{
        \subfigure
        {\includegraphics[width=3in]{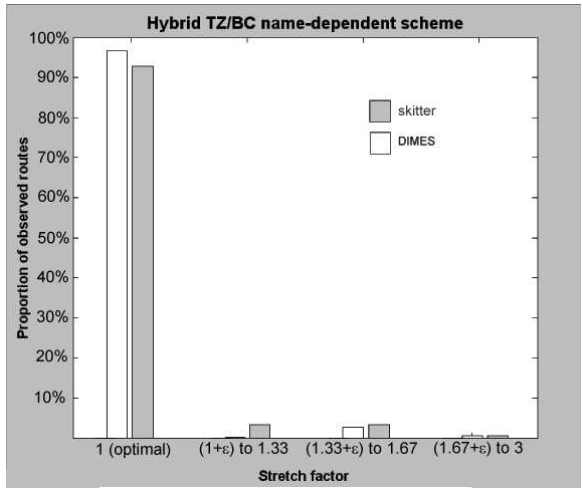}
        }
        \hfill
        \subfigure
        {\includegraphics[width=3in]{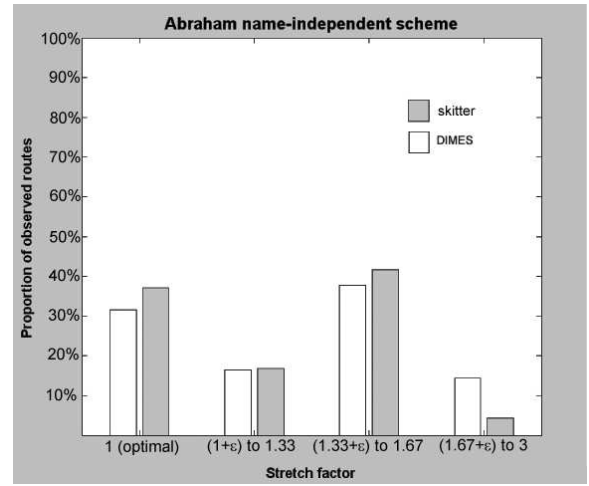}
        }
    }
    \centerline{
        \subfigure
        {\includegraphics[width=3in]{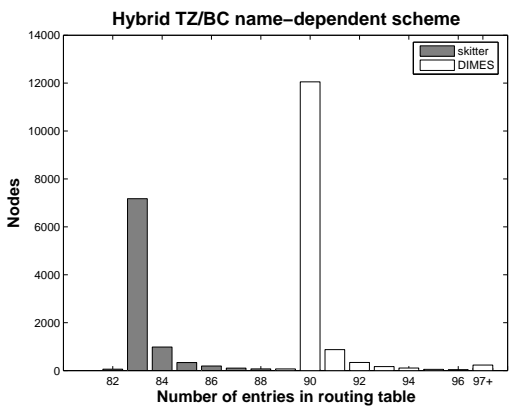}
        }
        \hfill
        \subfigure
        {\includegraphics[width=3in]{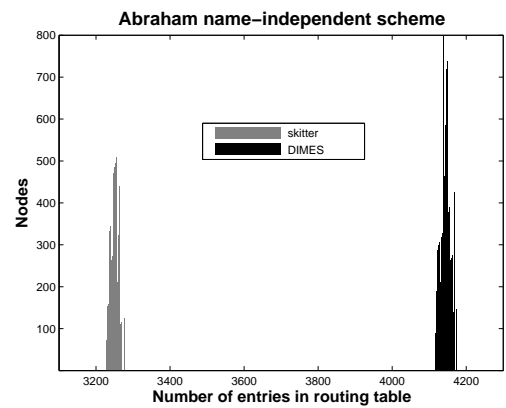}
        }
    }
    \caption{The distributions of stretch (top row) and routing table size (bottom row) that the hybrid
    TZ/BC (left column)
    and Abraham (right column) schemes produce on the skitter and DIMES topologies.
    \label{fig:distributions}
    }
\end{figure*}

To experimentally confirm this reasoning, we implemented the
best-performing name-dependent and name-independent schemes.
Specifically, among name-dependent schemes, we chose the universal
optimal stretch-3 TZ scheme~\cite{ThoZwi01b} and the BC
scheme~\cite{BraCow06} designed for scale-free graphs. We also
tested the TZ/BC hybrid scheme, which simply selects the
lower-stretch path by running both schemes in parallel. For the
name-independent case, we tested the optimal universal Abraham
scheme~\cite{AbGaMaNiTho04}. We then applied all schemes to AS-level
Internet topologies measured by skitter~\cite{as-adjacencies} and
DIMES~\cite{dimes}, and calculated the resulting average RT size and
stretch. Table~\ref{tab:tz-bc-abraham} and
Figure~\ref{fig:distributions} show the outcome of these
experiments, revealing that all characteristics of the
name-independent routing are significantly worse than their
name-dependent counterparts. By no means do these observations
conflict with our statements in Section~\ref{sect:compact_routing}
that the {\em worst-case\/} scaling of name-dependent and
name-independent universal routing schemes is the same. They do show,
however, that name-independent scaling on Internet topologies is much
worse {\em on average}.

\begin{table*}[t]
  \caption{\footnotesize The average stretch and routing table sizes (in entries
  and bits) that the best performing name-dependent (BC, TZ, and their hybrid TZ/BC) and
  name-independent (Abraham) schemes produce on the skitter and DIMES topologies.
  \label{tab:tz-bc-abraham}
  }
  \center
\begin{tabular}{|l|cc|cc|}
  \hline
  Scheme        & \multicolumn{2}{c|}{skitter (9204 nodes)} & \multicolumn{2}{c|}{DIMES (13931 nodes)} \\
  \hline
  TZ            & 62 entries   (1629 bits),   & 1.08 stretch      & 69 entries   (1887 bits),   & 1.13 stretch   \\
  BC            & 22 entries   (1025 bits),   & 1.06 stretch      & 22 entries   (1103 bits),   & 1.03 stretch   \\
  TZ/BC hybrid  & 84 entries   (2654 bits),   & 1.02 stretch      & 91 entries   (2990 bits),   & 1.01 stretch   \\
  Abraham       & 3251 entries (240427 bits), & 1.35 stretch      & 4145 entries (326781 bits), & 1.45 stretch   \\
  \hline
\end{tabular}
\end{table*}

Furthermore, our findings are consistent with the scaling properties
of name-independent routing on trees. We mention trees here because
tree routing from Section~\ref{sec:trees} forms the core of routing
schemes specialized for Internet topologies in
Section~\ref{sect:applicability}. Unfortunately, as found
in~\cite{LaRa05,AbGaMa06}, name-independent routing on trees cannot
scale better than on general graphs. In other words, {\em trees are
the worst-case graphs for universal name-independent
routing}.\footnote{More specifically, Abraham {\it et
al.}~\cite{AbGaMa06} explicitly use stars to prove the
name-independent lower bounds.} The reason for that is simple: trees
cannot guarantee that nodes' neighborhoods, i.e., balls of small
radii that store dictionary tables, comprise only small portions of
the graph. Examples of graphs that satisfy this property are grids
or their perturbations from Section~\ref{sec:grids}, i.e., graphs of
low doubling dimension. The average distance in these graphs grows
quickly with the graph size, the relative size of nodes'
neighborhoods can thus be made small, and as a consequence, the RT
sizes of name-independent routing scale logarithmically on such
graphs~\cite{AbGa06}. We summarize these scaling characteristics in
Table~\ref{tab:trees-grids}.

\begin{table*}[t]
  \caption{\footnotesize Routing table (RT) size scaling for name-dependent and
  name-independent routing on different graph classes. Both
  ``Logarithmic'' and ``Polynomial'' tags in this table have
  specific meanings. ``Logarithmic'' means that RT sizes scale as
  $\Omega(\log n)$ even for shortest-path (stretch-1) routing.
  ``Polynomial'' means $\Omega(n)$ for stretch-1 routing,
  $\Omega(\sqrt{n})$ for stretch-3 routing, and more generally,
  $\Omega(n^{1/k})$ for routing with stretch of~\mbox{$2k-1$}, \mbox{$k=1,2,\ldots$}.
  \label{tab:trees-grids}
  }
  \center
\begin{tabular}{|l|l|l|}
  \hline
  Graph class        & Name-dependent routing & Name-independent routing \\
  \hline
  General graphs     & Polynomial             & Polynomial  \\
  Trees              & Logarithmic            & Polynomial  \\
  Grids              & Logarithmic            & Logarithmic \\
  \hline
\end{tabular}
\end{table*}

\section{Conclusion}
\label{sect:conlcusion}

Compact routing is a research area that aims at solving the
following two classes of problems: 1)~identify the fundamental
scaling limits (lower bounds) of routing on graphs; and 2)~construct
routing algorithms that meet those limits, i.e., whose upper bounds
are equal to the corresponding lower bounds. In other words, compact
routing reveals the most fundamental properties of routing
scalability, refined from layers of complexity associated with
routing in practice.

As such, compact routing algorithms offer remarkably better scaling
than we observe in the Internet routing today. In particular,
existing interdomain routing exhibits exponentially growing routing
table sizes, while those of compact routing algorithms designed for
Internet-like topologies scale logarithmically.

Logarithmic scaling is essentially the only scaling behavior that
satisfies the requirement for a future routing architecture to
``scale indefinitely''~\cite{routing-requirements}. Unfortunately,
only static and name-dependent, i.e., topology-aware, routing
exhibit logarithmic scaling on observed Internet topologies.

If name-independence is a requirement, i.e., if we need to route on
topology-unaware flat identifiers, then routing tables cannot scale
better than polynomially on Internet-like topologies. The lower
bound is $\Omega(n)$ for shortest-path routing, $\Omega(\sqrt{n})$
for stretch-3 routing, and more generally, $\Omega(n^{1/k})$ for
routing with stretch of~\mbox{$2k-1$}, \mbox{$k=1,2,\ldots$}.

Worse than that, the communication cost, i.e., the number of routing
control messages per topology change, cannot grow slower than
linearly on Internet-like topologies. Of course, this scaling is
still much better than the exponential communication costs of
deployed routing protocols, but linear scaling can definitely not be
considered as satisfying the requirement of ``infinite
scalability.''

We are at an apparent impasse. Routing schemes specialized for
static Internet-like graphs scale essentially indefinitely, but in
adding the associated updates required to handle dynamic graphs, we
can no longer guarantee the required scaling behavior. We must face
the reality that scalable routing which uses topology update
messages to dynamically react to topology changes is not possible in
principle, which explains, in part, the lack of significant progress
in search of truly scalable routing.
We thus conclude that in order to find approaches that would lead us
to required routing scalability, we need some radically new ideas
that would allow us to construct convergence-free, ``updateless''
routing requiring no full view of network topologies. It is widely
known and popularized that having no full view of social network
topologies, humans can still efficiently route messages through
them~\cite{milgram67}. Whether we can devise routing protocols that
would do the same for the Internet is an open question.

\section{Acknowledgements}

This work is supported in part by NSF CNS-0434996.

\appendix

We demonstrate that the lower bound for communication costs in
random scale-free graphs of size~$n$ is~$\tilde{\Omega}(n)$. The
average distance~$\bar{d}$ in such graphs grows logarithmically
with~$n$, \mbox{$\bar{d} \sim \log n$}~\cite{ChLu03,DoMeSa03a},
while the width (standard deviation) of the distance distribution
approaches zero in the large graph limit~\cite{DoMeSa03a}.  The
latter property implies that the distance distribution~$d_x(r)$ for
every node~$x$, i.e., the number of nodes located at distance~$r$
from~$x$ divided by~$n$, approaches a delta-function as~\mbox{$n \to
\infty$}:
\begin{equation}\label{eq:dxr}
    d_x(r) \xrightarrow[n\to\infty]{}
    \begin{cases}
        1 & \text{if \; $r = \bar{d} \sim \log n$,}\\
        0 & \text{otherwise.}
    \end{cases}
\end{equation}
Korman and Peleg~\cite{KoPe06} define the {\em local density\/} as
\begin{equation}\label{eq:D}
D = \max_{x,r} \frac{|B_x(r)|}{2r},
\end{equation}
where $B_x(r)$ is the ball of radius~$r$ around node~$x$, i.e., the
set of nodes located within distance at most~$r$ from~$x$, and
$|B_x(r)|$ is the size of this set.  The maximum is taken over all
nodes~$x$ in the graph and all radii~$r$.  Observing
that~\mbox{$|B_x(r)| = n \int_0^r d_x(\rho)d\rho$} and using
eq.~(\ref{eq:dxr}), we immediately obtain that $|B_x(r)|$ in large
scale-free graphs is given by the Heaviside function:
\begin{equation}
    |B_x(r)| \xrightarrow[n\to\infty]{}
    \begin{cases}
        n & \text{if \; $r \geq \bar{d} \sim \log n$,}\\
        0 & \text{otherwise.}
    \end{cases}
\end{equation}
Substitution of this expression into eq.~(\ref{eq:D}) yields
\begin{equation}
D \xrightarrow[n\to\infty]{} \frac{n}{\log n}.
\end{equation}
The last expression and the demonstration in~\cite{KoPe06} that the
communication cost lower bound for graphs with local density~$D$
is~$\Omega(D)$ complete the proof.


\begin{thebibliography}{10}

\bibitem{iab-raws-report}
D.~Meyer, L.~Zhang, and {K. Fall (Eds.)}.
\newblock Report from the {IAB} workshop on routing and addressing.
\newblock IRTF, Internet Draft, 2007.
\newblock
  \url{http://www.ietf.org/internet-drafts/draft-iab-raws-report-02.txt}.

\bibitem{routing-requirements}
A.~Doria, E.~Davies, and {F. Kastenholz (Edts.)}.
\newblock Requirements for inter-domain routing.
\newblock IRTF, Internet Draft, 2006.

\bibitem{LaAhBo01}
C.~Labovitz, A.~Ahuja, A.~Bose, and F.~Jahanian.
\newblock Delayed {Internet} routing convergence.
\newblock {\em Transactions on Networking}, 9(3), 2001.

\bibitem{huston-bgp-site}
{BGP Table Data}, {\tt http://bgp.potaroo.net/}.

\bibitem{huston01-03}
G.~Huston.
\newblock Scaling inter-domain routing---a view forward.
\newblock {\em The Internet Protocol Journal}, 4(4), 2001.

\bibitem{huston01-02}
G.~Huston.
\newblock Commentary on inter-domain routing in the {Internet}.
\newblock IETF, RFC 3221, 2001.

\bibitem{BroNeCla02}
A.~Broido, E.~Nemeth, and kc~claffy.
\newblock Internet expansion, refinement, and churn.
\newblock {\em European Transactions on Telecommunications}, 13(1):33--51,
  2002.

\bibitem{NaGoVa03}
H.~Narayan, R.~Govindan, and G.~Varghese.
\newblock The impact of address allocation and routing on the structure and
  implementation of routing tables.
\newblock In {\em SIGCOMM}, 2003.

\bibitem{beyond-bgp}
Beyond {BGP}.
\newblock \url{http://www.beyondbgp.net/}.

\bibitem{atoms}
P.~Verkaik, A.~Broido, kc~claffy, R.~Gao, Y.~Hyun, and R.~van~der Pol.
\newblock Beyond {CIDR} aggregation.
\newblock Technical Report TR-2004-1, CAIDA, 2004.

\bibitem{multi6}
{Internet Engineering Task Force (IETF)}.
\newblock Site multihoming in {IPv6}.
\newblock Working Group.

\bibitem{islay}
F.~Kastenholz.
\newblock {ISLAY}: A new routing and addressing architecture.
\newblock IRTF, Internet Draft, 2002.

\bibitem{nimrod}
I.~Castineyra, N.~Chiappa, and M.~Steenstrup.
\newblock The nimrod routing architecture.
\newblock IETF, RFC 1992, 1996.

\bibitem{GuKoKi04}
R.~Gummadi, R.~Govindan, N.~Kothari, B.~Karp, Y.-J. Kim, and S.~Shenker.
\newblock Reduced state routing in the {Internet}.
\newblock In {\em HotNets}, 2004.

\bibitem{SuKaEe05}
L.~Subramanian, M.~Caesar, C.~T. Ee, M.~Handley, M.~Mao, S.~Shenker, and
  I.~Stoica.
\newblock {HLP}: A next generation inter-domain routing protocol.
\newblock In {\em SIGCOMM}, 2005.

\bibitem{CaCo06}
M.~Caesar, T.~Condie, J.~Kannan, K.~Lakshminarayanan, I.~Stoica, and
  S.~Shenker.
\newblock {ROFL}: Routing on flat labels.
\newblock In {\em SIGCOMM}, 2006.

\bibitem{GaPe96}
C.~Gavoille and S.~P\'{e}renn\`{e}s.
\newblock Memory requirement for routing in distributed networks.
\newblock In {\em PODC}, 1996.

\bibitem{SprMaAn03}
N.~Spring, R.~Mahajan, and T.~Anderson.
\newblock Quantifying the causes of path inflation.
\newblock In {\em SIGCOMM}, 2003.

\bibitem{BeGa04}
J.~Behrens and J.~J. Garcia-Luna-Aceves.
\newblock Distributed, scalable routing based on link-state vectors.
\newblock In {\em SIGCOMM}, 1994.

\bibitem{KK77}
L.~Kleinrock and F.~Kamoun.
\newblock Hierarchical routing for large networks: Performance evaluation and
  optimization.
\newblock {\em Computer Networks}, 1:155--174, 1977.

\bibitem{AbGa06}
I.~Abraham, C.~Gavoille, A.~Goldberg, and D.~Malkhi.
\newblock Routing in networks with low doubling dimension.
\newblock In {\em ICDCS}, 2006.

\bibitem{InMa04}
P.~Indyk and J.~Matou\v{s}ek.
\newblock {\em Handbook of Discrete and Computational Geometry}, chapter 8,
  Low-Distortion Embeddings of Finite Metric Spaces.
\newblock Chapman \& Hall/CRC, Boca Raton, 2004.

\bibitem{AbBa05}
I.~Abraham, Y.~Bartal, T.-H.~H. Chan, K.~Dhamdhere, A.~Gupta, J.~Kleinberg,
  O.~Neiman, and A.~Slivkins.
\newblock Metric embeddings with relaxed guarantees.
\newblock In {\em FOCS}, 2005.

\bibitem{FraGa01}
P.~Fraigniaud and C.~Gavoille.
\newblock Routing in trees.
\newblock In {\em ICALP}, 2001.

\bibitem{ThoZwi01b}
M.~Thorup and U.~Zwick.
\newblock Compact routing schemes.
\newblock In {\em SPAA}, 2001.

\bibitem{matousek02}
J.~Matou\v{s}ek.
\newblock {\em Lectures on Discrete Geometry}, chapter 15, Embedding Finite
  Metric Spaces into Normed Spaces.
\newblock Springer, New York, 2002.

\bibitem{KraLe06}
R.~Krauthgamer and J.~R. Lee.
\newblock Algorithms on negatively curved spaces.
\newblock In {\em FOCS}, 2006.

\bibitem{kleinberg07infocom}
R.~Kleinberg.
\newblock Geographic routing using hyperbolic space.
\newblock In {\em INFOCOM}, 2007.

\bibitem{MaKrFo06}
P.~Mahadevan, D.~Krioukov, M.~Fomenkov, B.~Huffaker, X.~Dimitropoulos,
  kc~claffy, and A.~Vahdat.
\newblock The {Internet} {AS}-level topology: Three data sources and one
  definitive metric.
\newblock {\em Computer Communication Review}, 36(1), 2006.

\bibitem{ChLu03}
F.~Chung and L.~Lu.
\newblock The average distance in a random graph with given expected degrees.
\newblock {\em Internet Mathematics}, 1(1):91--114, 2003.

\bibitem{LeCleFa05}
J.~Leskovec, J.~Kleinberg, and C.~Faloutsos.
\newblock Graphs over time: Densification laws, shrinking diameters and
  possible explanations.
\newblock In {\em KDD}, 2005.

\bibitem{KrFaYa04}
D.~Krioukov, K.~Fall, and X.~Yang.
\newblock Compact routing on {Internet}-like graphs.
\newblock In {\em INFOCOM}, 2004.

\bibitem{GaGe01}
C.~Gavoille and M.~Genegler.
\newblock Space-efficiency for routing schemes of stretch factor three.
\newblock {\em Journal of Parallel and Distributed Computing}, 61(5):679--687,
  2001.

\bibitem{ThoZwi01a}
M.~Thorup and U.~Zwick.
\newblock Approximate distance oracles.
\newblock In {\em STOC}, 2001.

\bibitem{cowen01}
L.~Cowen.
\newblock Compact routing with minimum stretch.
\newblock {\em Journal of Algorithms}, 38(1):170--183, 2001.

\bibitem{AbGaMaNiTho04}
I.~Abraham, C.~Gavoille, D.~Malkhi, N.~Nisan, and M.~Thorup.
\newblock Compact name-independent routing with minimum stretch.
\newblock In {\em SPAA}, 2004.

\bibitem{ArCoLaRaTa03}
M.~Arias, L.~Cowen, K.~A. Laing, R.~Rajaraman, and O.~Taka.
\newblock Compact routing with name independence.
\newblock In {\em SPAA}, 2003.

\bibitem{BraCow06}
A.~Brady and L.~Cowen.
\newblock Compact routing on power-law graphs with additive stretch.
\newblock In {\em ALENEX}, 2006.

\bibitem{CaCoDo06}
S.~Carmi, R.~Cohen, and D.~Dolev.
\newblock Searching complex networks efficiently with minimal information.
\newblock {\em Europhysics Letters}, 74:1102--1108, 2006.

\bibitem{GriWil99}
T.~Griffin and G.~Wilfong.
\newblock An analysis of {BGP} convergence properties.
\newblock In {\em SIGCOMM}, 1999.

\bibitem{GriWil02}
T.~Griffin and G.~Wilfong.
\newblock Analysis of the {MED} oscillation problem in {BGP}.
\newblock In {\em ICNP}, 2002.

\bibitem{AfGaRi89}
Y.~Afek, E.~Gafni, and M.~Ricklin.
\newblock Upper and lower bounds for routing schemes in dynamic networks.
\newblock In {\em FOCS}, 1989.

\bibitem{KoPe06}
A.~Korman and D.~Peleg.
\newblock Dynamic routing schemes for general graphs.
\newblock In {\em ICALP}, 2006.

\bibitem{as-adjacencies}
{CAIDA}.
\newblock Macroscopic topology {AS} adjacencies.
\newblock
  \url{http://www.caida.org/tools/measurement/skitter/as_adjacencies.xml}.

\bibitem{dimes}
The {DIMES} project.
\newblock \url{http://www.netdimes.org/}.

\bibitem{LaRa05}
K.~A. Laing and R.~Rajaraman.
\newblock A space lower bound for name-independent compact routing in trees.
\newblock In {\em SPAA}, 2005.

\bibitem{AbGaMa06}
I.~Abraham, C.~Gavoille, and D.~Malkhi.
\newblock On space-stretch trade-offs: Lower bounds.
\newblock In {\em SPAA}, 2006.

\bibitem{milgram67}
S.~Milgram.
\newblock The small world problem.
\newblock {\em Psychology Today}, 1:61--67, 1967.

\bibitem{DoMeSa03a}
S.~N. Dorogovtsev, J.~F.~F. Mendes, and A.~N. Samukhin.
\newblock Metric structure of random networks.
\newblock {\em Nuclear Physics B}, 653(3):307--422, 2003.

\end{thebibliography}
\end{document}